\documentclass[11pt]{article}
\usepackage[utf8]{inputenc}
\usepackage{amsmath,amssymb,amsthm}
\usepackage{graphicx}
\usepackage{hyperref}
\usepackage{natbib}
\usepackage{geometry}
\title{Residual Symmetry Reductions and Painlev\'e Solitons}
\author{
Li Yan$^{1}$, Xia Ya-Rong$^{2}$, Yao Ruo-Xia$^{1}$, Lou S. Y.$^{3,4}$ \\
\small{$^{1}$\textit{School of Artificial Intelligence and Computer Science, Shaanxi Normal University, Xian, 710119, China}} \\
\small{$^{2}$\textit{School of Information and Engineering, Xian University, Xian, Shaanxi 710065, China}} \\
\small{$^{3}$\textit{School of Physical Science and Technology, Ningbo University, Ningbo, 315211, China}} \\
\small{$^{4}$\textit{Institute of Fundamental Physics and Quantum Technology, Ningbo University, Ningbo, 315211, China}}
}
\date{}

\begin{document}

\maketitle
\begin{abstract}
This letter introduces the novel concept of {Painlev\'e solitons}---waves arising from the interaction between Painlev\'e waves and solitons in integrable systems. Painlev\'e solitons may also be viewed as solitons propagating against a Painlev\'e wave background, in analogy with the established notion of {elliptic solitons}, which refer to solitons on an elliptic wave background. By employing a novel symmetry decomposition method aided by nonlocal residual symmetries, we explicitly construct {(extended) Painlev\'e II solitons} for the Korteweg-de Vries (KdV) equation and {(extended) Painlev\'e IV solitons} for the Boussinesq equation.
\end{abstract}
The study of integrable nonlinear evolution equations has been a cornerstone of mathematical physics for decades, yielding profound insights into the behavior of wave phenomena across diverse fields. Among the most celebrated discoveries in this area is the {soliton}, a wave packet that maintains its shape and speed upon collision with other solitons, exhibiting particle-like properties \cite{zakharov1971, ablowitz1991}. The robustness and universality of solitons have made them a central object of study in fields as varied as fluid dynamics, nonlinear optics, plasma physics, and Bose-Einstein condensates, among others.

Parallel to the theory of solitons, the {Painlev\'e analysis} has served as a powerful integrability detector and solution-generating technique. The Painlev\'e property, which requires that all movable singularities of the solution are poles, provides a stringent test for integrability \cite{ablowitz1980, weiss1983}. Beyond mere classification, the Painlev\'e expansion method, particularly in its truncated form, offers a systematic algorithm for deriving exact solutions, including solitary waves and {Painlev\'e waves}. These Painlev\'e waves are solutions expressible in terms of the Painlev\'e transcendents, which are nonlinear special functions satisfying the six Painlev\'e equations (PI--PVI) \cite{clarkson2003}. They often describe rich, non-periodic oscillatory structures and act as important asymptotic backgrounds in many physical contexts.

A natural and fruitful generalization of the classical soliton concept involves placing it on a non-trivial background. A quintessential example is the {elliptic soliton}, which is defined as a soliton propagating on a background of elliptic (cnoidal) waves. These solutions capture the intricate interaction between a localized pulse and a periodic structure, leading to a more complex and physically relevant wave profile than a soliton in a vacuum \cite{smirnov2002}. The term ``elliptic soliton" elegantly encapsulates this hybrid nature, distinguishing it from its simpler counterpart. In Ref. \cite{ling2023}, Ling and Sun constructed multiple elliptic solitons and analyze their asymptotic behaviors for the focusing modified KdV equation via Darboux--B\"acklund transformation. Recently Nijhoff, Sun and Zhang \cite{nijhoff2023} introduced the concept of elliptic \(N\)th root of unity and established an elliptic direct linearization scheme (of Dutch version). Elliptic \(N\)th root of unity can be used to defined plane wave factors for the Boussinesq equations and the equations in the Gal'fand-Dicky hierarchy \cite{Dickey2003}. Meanwhile, a bilinear framework serving for elliptic solitons (based on the Lam\'e function) was established by Li and Zhang for both continuous \cite{li2022} and discrete systems. Vertex operators that generate \(\tau\) functions of elliptic solitons were found, and bilinear identities for the related \(\tau\) functions were given. More recently, an elliptic direct linearization scheme (Fokas--Ablowitz's version) was established by Li, Sun and Zhang \cite{li2025} for the KP, KdV and Boussinesq equations.

In recent years, the role of {nonlocal symmetries} has emerged as a pivotal tool for understanding and solving integrable systems. Unlike local symmetries, which depend on a finite number of derivatives of the field variables, nonlocal symmetries involve integrals of the dependent variables and are often connected to the recursion operator and B\"acklund transformations and Darboux transformations \cite{olver1993, bluman1989, lou1997, hu2009}. The concept of {residual symmetry}, which arises from the M\"obius invariance of the Painlev\'e expansion and remains after the truncation process, is a specific type of nonlocal symmetry that has proven particularly effective. This residual symmetry can be localized by introducing a suitable auxiliary system, thereby enabling the application of standard Lie group methods to generate new and complex solutions from known ones \cite{gao2013, guo2012, liu2018}.

Building upon these foundations, this letter introduces and develops the novel concept of {Painlev\'e solitons}. We define these as sophisticated wave structures that represent the nonlinear interaction between a classical soliton and a Painlev\'e wave. In essence, a Painlev\'e soliton can be conceptualized as a soliton riding upon a dynamic, non-periodic background governed by a Painlev\'e transcendent. This definition is in direct analogy with the established notion of elliptic solitons, where the elliptic wave background is replaced by the more complex Painlev\'e wave background. The nomenclature {Painlev\'e soliton} is thus proposed to describe this new class of solutions, highlighting their composite nature.

The primary objective of this work is to formally establish the existence of Painlev\'e solitons and to provide a constructive method for obtaining them. To this end, we employ a novel {symmetry decomposition method} that leverages the power of nonlocal residual symmetries. This methodology allows us to systematically deconstruct the solution process, isolating the contributions from the soliton and the Painlev\'e background. Through this approach, we successfully derive explicit forms of two distinct types of Painlev\'e solitons:
\begin{itemize}
    \item \textbf{Painlev\'e II solitons} for the Korteweg-de Vries (KdV) equation, and
    \item \textbf{Painlev\'e IV solitons} for the Boussinesq equation.
\end{itemize}

The KdV equation, a paradigmatic model for shallow water waves, and the Boussinesq equation, which describes waves in dispersive media, provide excellent testbeds for our theory. The derivation of Painlev\'e II and IV solitons for these systems not only validates our conceptual framework but also demonstrates the efficacy of the symmetry decomposition method.

It is known that the KdV equation,
\begin{equation}
u_{t} = u_{xxx} + 6uu_{x},
\label{eq:kdv}
\end{equation}
possesses a symmetry \cite{liu2018}
\begin{equation}
\sigma = \frac{1}{3}c\lambda-2cu-bh + \frac{1}{6}\delta-(cx - 2c\lambda t - \delta t + x_{0})u_{x} - (3ct + t_{0})u_{t},
\label{eq:symmetry}
\end{equation}
which is a solution of the linearized equation
\begin{equation}
\sigma_{t} = \sigma_{xxx} + 6\sigma u_{x} + 6u\sigma_{x}
\label{eq:linearized_kdv}
\end{equation}
of the KdV equation (\ref{eq:kdv}), where
\begin{equation}
h = f_{xx}
\label{eq:h_def}
\end{equation}
and \(f\) is a solution of the Schwartz KdV equation
\begin{equation}
f_{t} = h_{x} - \frac{3}{2}h^{2}g^{-1} + \lambda g
\label{eq:schwartz_kdv}
\end{equation}
with
\begin{equation}
g = f_{x},
\label{eq:g_def}
\end{equation}
while the KdV equation (\ref{eq:kdv}) and the Schwartz KdV equation (\ref{eq:schwartz_kdv}) are related by nonauto-B\"acklund transformation
\begin{equation}
u = -\frac{2}{3}\frac{h_{x}}{g} + \frac{1}{6}\frac{f_{t}}{g} + \frac{1}{2}\frac{h^{2}}{g^{2}}
\label{eq:backlund}
\end{equation}
and/or the truncated Painlev\'e expansion \(u = -2\frac{g^{2}}{f^{2}} + 2\frac{h}{f} - \frac{2}{3}\frac{h_{x}}{g} + \frac{1}{6}\frac{f_{t}}{g} + \frac{1}{2}\frac{h^{2}}{g^{2}}\).

In the symmetry expression (\ref{eq:symmetry}), the constant $c$ related parts constitute the local scaling symmetry, $x_0$ and $t_0$ parts are related to the local space time translation invariance,  $\delta$ parts correspond to the local Galileo transformation, $\lambda$ is same as in the Schwartz KdV equation (\ref{eq:schwartz_kdv}) and also related to the spectral parameter.  The term $bh$ of (\ref{eq:symmetry}) is a nonlocal symmetry of the single KdV equation (\ref{eq:kdv}). However, (\ref{eq:symmetry}) is a local symmetry of the prolonged system (\ref{eq:kdv}), (\ref{eq:h_def}), (\ref{eq:g_def}) and (\ref{eq:backlund}) for the fields \(u, f, g\) and \(h\).

The full symmetry equation system of (\ref{eq:kdv}), (\ref{eq:h_def}), (\ref{eq:g_def}) and (\ref{eq:backlund}) can be written as (\ref{eq:linearized_kdv}),
\begin{align}
\sigma_{3} &= \sigma_{1xx}, \label{eq:sym_system1} \\
\sigma_{2} &= \sigma_{1x} \label{eq:sym_system2}
\end{align}
and
\begin{equation}
\sigma = -\frac{2}{3}\frac{\sigma_{3x}}{g} + \frac{2}{3}\frac{h_{x}\sigma_{2}}{g^{2}} + \frac{1}{6}\frac{\sigma_{1t}}{g} - \frac{1}{6}\frac{f_{t}\sigma_{2}}{g^{2}} + \frac{h\sigma_{3}}{g^{2}} - \frac{h^{2}\sigma_{2}}{g^{3}}.
\label{eq:sym_system3}
\end{equation}

Related to the symmetry (\ref{eq:symmetry}), the corresponding symmetry solutions of (\ref{eq:sym_system1}), (\ref{eq:sym_system2}) and (\ref{eq:sym_system3}) read
\begin{align}
\sigma_{1} &= \frac{1}{2}bf^{2} + df + e - (cx - 2c\lambda t - \delta t + x_{0})f_{x} - (3ct + t_{0})f_{t}, \label{eq:sym_sol1} \\
\sigma_{2} &= g(bf + d - c) - (cx - 2c\lambda t - \delta t + x_{0})g_{x} - (3ct + t_{0})g_{t}, \label{eq:sym_sol2} \\
\sigma_{3} &= bg^{2} + h(bf + d - 2c) - (cx - 2c\lambda t - \delta t + x_{0})h_{x} - (3ct + t_{0})h_{t}. \label{eq:sym_sol3}
\end{align}

To solve the group invariant solutions related to the symmetries (\ref{eq:symmetry}), (\ref{eq:sym_sol1}), (\ref{eq:sym_sol2}) and (\ref{eq:sym_sol3}), one can discuss two subcases \(c \neq 0\) and \(c = 0\).\\
\textbf{Case 1: \(c = 0\).}
In this case, without loss of generality, one can take \(t_{0} = -1\) and the group invariant conditions \(\sigma = \sigma_{1} = \sigma_{2} = \sigma_{3} = 0\) lead to the solutions of (\ref{eq:symmetry}), (\ref{eq:sym_sol1}), (\ref{eq:sym_sol2}) and (\ref{eq:sym_sol3}) in the forms
\begin{equation}
u = U(\xi, t), \quad f = F(\xi, t), \quad g = G(\xi, t), \quad h = H(\xi, t), \quad \xi = x + x_{0}t
\label{eq:case1_vars}
\end{equation}
and
\begin{align}
U_{t} &= bH, \label{eq:case1_system1} \\
F_{t} &= -\frac{1}{2}bF^{2} - dF - e, \label{eq:case1_system2} \\
G_{t} &= -G(bF + d), \label{eq:case1_system3} \\
H_{t} &= -H(bF + d) - G^{2}. \label{eq:case1_system4}
\end{align}

Substituting the relations (\ref{eq:case1_vars}) and (\ref{eq:case1_system1}--\ref{eq:case1_system4}) into the prolonged KdV system (\ref{eq:kdv}), (\ref{eq:h_def}), (\ref{eq:g_def}) and (\ref{eq:backlund}), we have
\begin{align}
F_{\xi} &= G, \label{eq:xi_system1} \\
G_{\xi} &= H, \label{eq:xi_system2} \\
H_{\xi} &= -2UG + \frac{1}{3}\lambda G + \frac{H^{2}}{2G}, \label{eq:xi_system3} \\
U_{\xi\xi\xi} &= (x_{0} - 6U)U_{\xi} + bH. \label{eq:xi_system4}
\end{align}

One can directly verify that the finite dimensional dynamic systems (\ref{eq:case1_system1}--\ref{eq:case1_system4}) and (\ref{eq:xi_system1}--\ref{eq:xi_system4}) are consistent because \(F_{\xi t} = F_{t\xi}\), \(G_{\xi t} = G_{t\xi}\), \(H_{\xi t} = H_{t\xi}\) and \(U_{\xi\xi\xi t} = U_{t\xi\xi\xi}\) are naturally satisfied.

From the consistent dynamic systems (\ref{eq:case1_system1}--\ref{eq:case1_system4}) and (\ref{eq:xi_system1}--\ref{eq:xi_system4}), we find that the (1+1)-dimensional integrable system (\ref{eq:kdv}), (\ref{eq:h_def}), (\ref{eq:g_def}) and (\ref{eq:backlund}) can be solved by a \(t\)-dynamic system (\ref{eq:case1_system1}--\ref{eq:case1_system4}) and a \(\xi\)-dynamic system (\ref{eq:xi_system1}--\ref{eq:xi_system4}). This kind of decomposition method appeared in two different situations. The first one is the so-called nonlinearization approach \cite{cao1990, cheng1991} for the Lax pair. The second one is the so-called formal variable separation approach \cite{hao2022, lou1999}. The decomposition method proposed in \cite{hao2022} can solve high dimensional systems via consistent lower dimensional ones. On the other hand, lower dimensional integrable systems can be combined to higher dimensional integrable ones. Furthermore, the decomposition idea yields a deformation algorithm \cite{Lou2023} to find new higher dimensional integrable systems and new types of solitons--twisted nontravelling solitary waves \cite{Lou2025}.

The consistent dynamical systems can be simply solved with the final results (\(\xi \equiv x + x_{0}t\))
\begin{align}
u &= U(\xi) + \frac{b}{a}H(\xi)\tanh(a\phi) + \frac{b^{2}}{2a^{2}}G(\xi)^{2}\text{sech}^{2}(a\phi), \quad \phi \equiv t + F(\xi) \label{eq:sol1_u} \\
f &= -\frac{d}{b} + \frac{2a}{b}\tanh(a\phi), \label{eq:sol1_f} \\
g &= G(\xi)\text{sech}^{2}(a\phi), \label{eq:sol1_g} \\
h &= \frac{1}{a}\text{sech}^{2}(a\phi)\left[aH(\xi) - bG(\xi)^{2}\tanh(a\phi)\right], \label{eq:sol1_h}
\end{align}
where \(a \neq 0\) is related to nonlocal symmetry parameters by
\begin{equation}
a^{2} = \frac{1}{4}(d^{2} - 2be),
\end{equation}
while the group invariant functions \(F = F(\xi), G = G(\xi), H = H(\xi)\) and \(U = U(\xi)\) are determined by
\begin{align}
U &= -\frac{b^{2}}{2a^{2}}G^{2} + \frac{2}{3}\lambda - \frac{x_{0}}{2} - \frac{a^{2}}{bG} - \frac{H^{2}}{2G^{2}}, \label{eq:U_relation} \\
H &= G_{\xi}, \label{eq:H_relation} \\
F &= \frac{b}{2a^{2}}\int^{\xi}G(y)\mathrm{d}y \label{eq:F_relation}
\end{align}
and
\begin{equation}
G_{\xi}^{2} = \frac{b^{2}}{a^{2}}G^{4} + CG^{3} + 2(\lambda - x_{0})G^{2} - \frac{2a^{2}G}{b}.
\label{eq:G_equation}
\end{equation}

It is clear that the \(G\) function of (\ref{eq:G_equation}) is an elliptic integral which can be explicitly expressed by Jacobi or Weierstrass elliptic functions. Thus, the solution (\ref{eq:sol1_u}--\ref{eq:sol1_h}) is just the well known elliptic soliton of the KdV equation (\ref{eq:kdv}).

\textbf{Case 2: \(c \neq 0\).}
In this case, without loss of generality, we can take
\begin{equation}
c = 1, \quad x_{0} = t_{0} = 0, \quad \delta = -2\lambda - c_{0} = 0.
\label{eq:case2_constants}
\end{equation}

Under the constant conditions of (\ref{eq:case2_constants}), the constraint conditions \(\sigma = \sigma_{1} = \sigma_{2} = \sigma_{3} = 0\) become
\begin{equation}
u = U(\eta, t), \quad f = F(\eta, t), \quad g = G(\eta, t), \quad h = H(\eta, t), \quad \eta \equiv \frac{x + \lambda t}{t^{1/3}},
\label{eq:case2_vars}
\end{equation}
and
\begin{align}
tU_{t} &= \frac{1}{9}(3cH - 6U + \lambda), \label{eq:case2_system1} \\
tF_{t} &= -\frac{1}{6c}(cF - 6a - c_{1})(cF + 6a - c_{1}), \label{eq:case2_system2} \\
tG_{t} &= -\frac{1}{3}G(cF - c_{1} + 1), \label{eq:case2_system3} \\
tH_{t} &= -\frac{c}{3}G^{2} - \frac{1}{3}H(cF - c_{1} + 2). \label{eq:case2_system4}
\end{align}

Substituting the expressions (\ref{eq:case2_vars}) and (\ref{eq:case2_system1}--\ref{eq:case2_system4}) into the equation system (\ref{eq:kdv}), (\ref{eq:h_def}), (\ref{eq:g_def}) and (\ref{eq:backlund}), we get an \(\eta\)-dynamic system
\begin{align}
U &= \frac{1}{6}\eta t^{-2/3} + \frac{1}{6}\lambda + \frac{(cF - 6a - c_{1})(cF + 6a - c_{1})}{12ctG} - \frac{H^{2}}{2G^{2}}, \label{eq:eta_system1} \\
F_{\eta} &= Gt^{1/3}, \label{eq:eta_system2} \\
G_{\eta} &= Ht^{1/3}, \label{eq:eta_system3} \\
H_{\eta} &= \frac{3H^{2}t^{1/3}}{2G} - \frac{\eta G}{3t^{1/3}} - \frac{(cF - 6a - c_{1})(cF + 6a - c_{1})}{6ct^{2/3}}. \label{eq:eta_system4}
\end{align}

The group invariant solution of the consistent symmetry decomposition systems (\ref{eq:case2_system1}--\ref{eq:case2_system4}) and (\ref{eq:eta_system1}--\ref{eq:eta_system4}) possesses the form
\begin{align}
f &= \frac{c_{1}}{c} + \frac{6a}{c}\tanh(\psi) = \frac{c_{1}}{c} + \frac{6a}{c}\frac{t^{2a}\exp(2aF(\eta))- 1}{t^{2a}\exp(2aF(\eta)) + 1}, \quad \psi \equiv aF(\eta) + \ln(t), \label{eq:sol2_f} \\
g &= \frac{6a^{2}}{ct^{1/3}}G(\eta)\text{sech}^{2}(\psi), \label{eq:sol2_g} \\
h &= \frac{6a^{2}}{ct^{2/3}}\text{sech}^{2}(\psi)\left[H(\eta) - 2aG(\eta)^{2}\tanh(\psi)\right], \label{eq:sol2_h} \\
u &= \frac{\lambda}{6} + \frac{\eta}{6t^{2/3}} - \frac{1}{2t^{2/3}G(\eta)} - \frac{H(\eta)^{2}}{2t^{2/3}G(\eta)^{2}} + \frac{2a}{t^{2/3}}\tanh(\psi)\left[H(\eta) - aG(\eta)^{2}\tanh(\psi)\right] \label{eq:sol2_u}
\end{align}
with \(G(\eta) = F_{\eta}(\eta), H(\eta) = F_{\eta\eta}(\eta)\) while the \(F(\eta) = F\) satisfy
\begin{equation}
6F_{\eta}F_{\eta\eta\eta} - 9F_{\eta\eta}^{2} + 2\eta F_{\eta}^{2} - 6F_{\eta} - 12a^{2}F_{\eta}^{4} = 0.
\label{eq:F_equation}
\end{equation}

Thus, whence the solution of (\ref{eq:F_equation}) is fixed, the corresponding solution (\ref{eq:sol2_u}) of the KdV equation (\ref{eq:kdv}) is obtained. The solution (\ref{eq:sol2_u}) is a soliton solution under the \(F\) background.

To see the property of (\ref{eq:F_equation}), we make the following transformation (\(P(\zeta) = P\)),
\begin{equation}
F(\eta) = F_{1}(b\eta) = F_{1}(\zeta) = F_{1}, \quad F_{1\zeta} = \frac{3}{\zeta + 2P^{2} + 2\delta_{1}P_{\zeta}}, \quad \delta_{1}^{2} = 1.
\label{eq:P_transformation}
\end{equation}

Substituting the transformation (\ref{eq:P_transformation}) into (\ref{eq:F_equation}) yields
\begin{align}
P_{\zeta\zeta} &= 2P^{3} + \zeta P + 3a\delta_{2} - \frac{\delta_{1}}{2} - \frac{6a\delta_{2}}{1 + A\exp\left(-6a\delta_{2}Q\right)}, \label{eq:extended_PII} \\
Q_{\zeta} &= \frac{1}{\delta_{1}\zeta + 2\delta_{1}P^{2} + 2P_{\zeta}}, \quad \delta_{1}^{2} = \delta_{2}^{2} = 1 \label{eq:Q_equation}
\end{align}
with an arbitrary constant \(A\).

It is clear that when \(A = 0\), the \(P\)-equation (\ref{eq:extended_PII}) is just the standard Painlev\'e II equation. Thus the general \(P\)-equation (\ref{eq:extended_PII}) with \(A \neq 0\) can be called as the extended Painlev\'e II equation. It should be emphasized that the symmetry reduction (\ref{eq:F_equation}) or (\ref{eq:extended_PII}) is a new important Painlev\'e reduction because (\ref{eq:extended_PII}) is beyond of the form,
\begin{equation}
P_{\zeta\zeta} = W(\zeta, P, P_{\zeta}),
\label{eq:standard_form}
\end{equation}
where \(W(\zeta, P, P_{\zeta})\) is rational in \(P_{\zeta}\), algebraic in \(P\) and analytic in \(\zeta\).

Accordingly, the solution (\ref{eq:sol2_u}) is an extended Painlev\'e II soliton for \(A \neq 0\) and a Painlev\'e II soliton for \(A = 0\). In other words, it is a soliton ride on an extended Painlev\'e wave.

In summary, for the KdV equation (\ref{eq:kdv}), by means of the nonlocal residual symmetry and local symmetries, one can obtain some types of symmetry decompositions including the elliptic solitons, extended Painlev\'e II solitons and Painlev\'e II solitons. Though the KdV is well known and studied by various experts, there are still many mysteries including the firstly discovered extended Painlev\'e II equation (\ref{eq:extended_PII}) in this paper.

Similarly, there are various uncovered mysteries for all well known integrable systems. To support this opinion, we study the residual symmetry reductions of the Boussinesq equation,
\begin{equation}
u_{tt} = (u_{xx} + u^{2})_{xx}.
\label{eq:boussinesq}
\end{equation}

By using the truncated Painlev\'e expansion method, one can find that both the transformation
\begin{equation}
u = \frac{1}{2}\frac{f_{t}^{2}}{f_{x}^{2}} - 2\frac{f_{xxx}}{f_{x}} + \frac{3}{2}\frac{f_{xx}^{2}}{f_{x}^{2}}
\label{eq:boussinesq_transformation}
\end{equation}
and the truncated expansion
\begin{equation}
u = -6\frac{f_{x}^{2}}{f^{2}} + 6\frac{f_{xx}}{f} + \frac{1}{2}\frac{f_{t}^{2}}{f_{x}^{2}} - 2\frac{f_{xxx}}{f_{x}} + \frac{3}{2}\frac{f_{xx}^{2}}{f_{x}^{2}}
\label{eq:boussinesq_truncated}
\end{equation}
can transform the Boussinesq equation (\ref{eq:boussinesq}) to its Schwartz Boussinesq equation
\begin{equation}
R_{t} = S_{x} - RR_{x}, \quad R = \frac{f_{t}}{f_{x}}, \quad S = \frac{f_{xxx}}{f_{x}} - \frac{3}{2}\frac{f_{xx}^{2}}{f_{x}^{2}}.
\label{eq:schwartz_boussinesq}
\end{equation}

One can directly verify that if \(u\) and \(f\) are related by (\ref{eq:boussinesq_transformation}), the residual, \(f_{xx}\), with respect to the arbitrary singular manifold \(f\), is a nonlocal symmetry of the Boussinesq equation (\ref{eq:boussinesq}).

Similar to the KdV equation, though the residual symmetry \(f_{xx}\) is nonlocal for the Boussinesq equation (\ref{eq:boussinesq}), however, it can become a local symmetry for the extended Boussinesq system (\ref{eq:boussinesq}), (\ref{eq:h_def}), (\ref{eq:g_def}) and (\ref{eq:boussinesq_transformation}).

Using the standard classical symmetry approach, it is not difficult to find that the Lie point symmetry of the extended Boussinesq system (\ref{eq:boussinesq}), (\ref{eq:h_def}), (\ref{eq:g_def}) and (\ref{eq:boussinesq_transformation}) possesses the form
\begin{align}
\sigma &= ah - 2cu - (cx + x_{0})u_{x} - (2ct + t_{0})u_{t}, \label{eq:bouss_sym1} \\
\sigma_{1} &= c_{0} + c_{1}f - \frac{1}{6}af^{2} - (cx + x_{0})f_{x} - (2ct + t_{0})f_{t}, \label{eq:bouss_sym2} \\
\sigma_{2} &= g\left(c_{1} - c - \frac{1}{3}af\right) - (cx + x_{0})g_{x} - (2ct + t_{0})g_{t}, \label{eq:bouss_sym3} \\
\sigma_{3} &= h\left(c_{1} - \frac{1}{3}af\right) - \frac{1}{3}ag^{2} - (cx + x_{0})h_{x} - (2ct + t_{0})h_{t}. \label{eq:bouss_sym4}
\end{align}

To find group invariant solution related to the symmetry (\ref{eq:bouss_sym1}--\ref{eq:bouss_sym4}), we discuss two subcases:\\
\textbf{Case I:} \(c = 0, t_{0} = 1, c_{0} = \frac{3(a_{1}^{2} - c_{1}^{2})}{2a}\).
In this case, the group invariant condition, \(\sigma = \sigma_{1} = \sigma_{2} = \sigma_{3} = 0\) yields the \(t\)-dynamic system,
\begin{align}
U_{t} &= -aH, \quad u = U(x + x_{0}t, t) = U(\xi, t), \quad h = H(\xi, t), \quad g = G(\xi, t), \quad f = F(\xi, t), \label{eq:bouss_case1_vars} \\
F_{t} &= \frac{a}{6}F^{2} - c_{1}F + \frac{3}{2a}(c_{1}^{2} - a_{1}^{2}), \label{eq:bouss_case1_system1} \\
G_{t} &= \frac{a}{3}FG - c_{1}G, \label{eq:bouss_case1_system2} \\
H_{t} &= \frac{a}{3}G^{2} + \frac{a}{3}FH - c_{1}H. \label{eq:bouss_case1_system3}
\end{align}

Substituting the \(t\)-dynamic system (\ref{eq:bouss_case1_system1}--\ref{eq:bouss_case1_system3}) into the prolonged Boussinesq system (\ref{eq:boussinesq}), (\ref{eq:h_def}), (\ref{eq:g_def}) and (\ref{eq:boussinesq_transformation}), one can get the \(\xi\)-dynamic system
\begin{align}
F_{\xi} &= G, \label{eq:bouss_xi_system1} \\
G_{\xi} &= H, \label{eq:bouss_xi_system2} \\
H_{\xi} &= \frac{1}{4}(x_{0}^{2} - 2U)G + \left[\frac{a}{12}F^{2} - \frac{c_{1}}{2}F - \frac{3}{4a}(a_{1}^{2} - c_{1}^{2})\right]x_{0} \nonumber \\
&\quad + \frac{1}{G}\left[\frac{3}{4}H^{2} + \frac{a^{2}F^{4}}{144} - \frac{ac_{1}F^{3}}{12} + \frac{3c_{1}^{2} - a_{1}^{2}}{8}F^{2} + \frac{3c_{1}}{4a}(a_{1}^{2} - c_{1}^{2})F + \frac{(a_{1}^{2} - c_{1}^{2})^{2}}{16a^{2}}\right], \label{eq:bouss_xi_system3} \\
H_{\xi} &= \frac{U_{\xi\xi}}{a}\left(\frac{x_{0}}{2} - \frac{U}{x_{0}}\right) - \frac{U_{\xi\xi\xi\xi}}{2ax_{0}} + \frac{H}{6x_{0}}(3c_{1} - aF) - \frac{aG^{2}}{6x_{0}} - \frac{U_{\xi}^{2}}{ax_{0}}. \label{eq:bouss_xi_system4}
\end{align}

The symmetry decomposition system (\ref{eq:bouss_case1_system1}--\ref{eq:bouss_case1_system3}) and (\ref{eq:bouss_xi_system1}--\ref{eq:bouss_xi_system4}) possesses the generalized solution
\begin{align}
f &= \frac{3c_{1}}{a} - \frac{3a_{1}}{a}\tanh(\theta), \quad \theta \equiv \frac{a_{1}t}{2} + F(\xi), \quad \xi = x + x_{0}t, \label{eq:bouss_sol1_f} \\
g &= G\text{sech}^{2}(\theta), \label{eq:bouss_sol1_g} \\
h &= \frac{1}{3a_{1}}\text{sech}^{2}(\theta)\left[2aG^{2}\tanh(\theta) + 3a_{1}H\right], \label{eq:bouss_sol1_h} \\
u &= U(\xi) - \frac{2aH(\xi)}{a_{1}}\tanh(\theta) + \frac{2a^{2}G^{2}}{3a_{1}^{2}}\text{sech}^{2}(\theta), \label{eq:bouss_sol1_u}
\end{align}
where
\begin{align}
H(\xi) &= G_{\xi}(\xi), \label{eq:bouss_relations1} \\
G(\xi) &= -\frac{3a_{1}}{a}F_{\xi}(\xi), \label{eq:bouss_relations2} \\
U(\xi) &= \frac{3H(\xi)^{2}}{2G(\xi)^{2}} - \frac{2H_{\xi}(\xi)}{G(\xi)} - \frac{2a^{2}G(\xi)^{2}}{9a_{1}^{2}} + \frac{x_{0}^{2}}{2} - \frac{3a_{1}^{2}x_{0}}{2aG(\xi)} + \frac{9a_{1}^{4}}{a^{2}G(\xi)^{2}}, \label{eq:bouss_relations3}
\end{align}
while \(G(\xi)\) is given by the following elliptic integral
\begin{equation}
\int^{G(\xi)}\frac{\mathrm{d}\mu}{\sqrt{V(\mu)}} = \xi + \xi_{0}, \quad V(\mu) = \frac{4a^{2}\mu^{4}}{9a_{1}^{2}} - C_{2}\mu^{3} + C_{1}\mu^{2} + \frac{3a_{1}^{2}x_{0}}{a}\mu - \frac{3a_{1}^{4}}{4a^{2}} \label{eq:bouss_elliptic}
\end{equation}
with three more arbitrary integral constants \(C_{1}, C_{2}\) and \(\xi_{0}\). It is obvious that the solution (\ref{eq:bouss_sol1_u}) is just the elliptic soliton solution of the Boussinesq equation (\ref{eq:boussinesq}).

\textbf{Case II:} \(c \neq 0, x_{0} = t_{0} = 0\).
In this case, we take the transformation,
\begin{equation}
u = U(\eta, t), \quad f = F(\eta, t), \quad g = G(\eta, t), \quad h = H(\eta, t), \quad \eta = \frac{x}{\sqrt{t}}.
\label{eq:bouss_case2_vars}
\end{equation}

Using the transformation (\ref{eq:bouss_case2_vars}), the second type of symmetry decomposition possesses the form
\begin{align}
2ctF_{t} &= -\frac{1}{6}aF^{2} + c_{1}F + c_{0}, \label{eq:bouss_case2_system1} \\
2ctG_{t} &= -\frac{1}{3}aFG - cG + c_{1}G, \label{eq:bouss_case2_system2} \\
2ctH_{t} &= -\frac{1}{3}aFH - \frac{1}{3}aG^{2} + c_{1}H - 2cH, \label{eq:bouss_case2_system3} \\
2ctU_{t} &= -2cU + aH. \label{eq:bouss_case2_system4}
\end{align}
and
\begin{align}
F_{\eta} &= \sqrt{t}G, \label{eq:bouss_eta_system1} \\
G_{\eta} &= \sqrt{t}H, \label{eq:bouss_eta_system2} \\
H_{\eta} &= \frac{(\xi^{2} - 8U)G}{16\sqrt{t}} + \frac{\eta}{48tc}\left[aF^{2} - 6c_{1}F - 9(a_{1}^{2} - c_{1}^{2})a^{-1}\right] \nonumber \\
&\quad + \frac{1}{t^{3/2}a^{2}c^{2}G}\left[a^{4}F^{4} - 12c_{1}a^{3}F^{3} - 18a^{2}(a_{1}^{2} - 3c_{1}^{2})F^{2} + 108ac_{1}(a_{1}^{2} - c_{1}^{2})F \right. \nonumber \\
&\quad \left. + 432a^{2}c^{2}t^{2}H^{2} + 81(a_{1}^{2} - c_{1}^{2})^{2}\right], \label{eq:bouss_eta_system3} \\
U_{\eta} &= \frac{\eta H(F_{1}^{2} - 9a_{1}^{2})}{16catG^{2}} - \frac{\xi(5c + F_{1})}{4ct} + \frac{3H(8tU - \eta^{2})}{16\sqrt{t}G} - \frac{(F_{1}^{2} - 9a_{1}^{2})(F_{1} + 9c)}{72ac^{2}t^{3/2}G} \nonumber \\
&\quad + \frac{3\sqrt{t}H^{3}}{4G^{3}} + \frac{(F_{1}^{2} - 9a_{1}^{2})^{2}H}{576a^{2}c^{2}t^{3/2}}. \label{eq:bouss_eta_system4}
\end{align}

The general solution of the symmetry decomposition (\ref{eq:bouss_case2_system1}--\ref{eq:bouss_case2_system4}) and (\ref{eq:bouss_eta_system1}--\ref{eq:bouss_eta_system4}) reads
\begin{align}
f &= \frac{3c_{1}}{a} + \frac{3a_{1}}{a}\tanh(\Phi), \quad \Phi = \frac{a_{1}}{4c}\ln(t) + F(\eta), \label{eq:bouss_sol2_f} \\
g &= \frac{G(\eta)}{\sqrt{t}}\text{sech}^{2}(\Phi), \label{eq:bouss_sol2_g} \\
h &= \frac{H(\eta)}{t}\text{sech}^{2}(\Phi) - \frac{2aG(\eta)^{2}}{3a_{1}t}\text{sech}^{2}(\Phi)\tanh(\Phi), \label{eq:bouss_sol2_h} \\
u &= \frac{U(\eta)}{t} + \frac{2a^{2}G(\eta)^{2}}{3a_{1}^{2}t}\text{sech}^{2}(\Phi) + \frac{2aH(\eta)}{a_{1}t}\tanh(\Phi), \label{eq:bouss_sol2_u}
\end{align}
where the group invariant functions \(F(\eta) = F, G(\eta) = G, H(\eta) = H\) and \(U(\eta) = U\) are related by
\begin{align}
F &= \frac{a}{3a_{1}}\int G\mathrm{d}\eta, \label{eq:bouss_sol2_relations1} \\
H &= G_{\eta}, \label{eq:bouss_sol2_relations2} \\
U &= \frac{3G_{\eta}^{2}}{2G^{2}} - \frac{2G_{\eta\eta}}{G} - \frac{2a^{2}G^{2}}{9a_{1}^{2}} + \frac{\eta^{2}}{8} - \frac{3a_{1}^{2}\eta}{8acG} + \frac{9a_{1}^{4}}{32a^{2}c^{2}G^{2}}, \label{eq:bouss_sol2_relations3}
\end{align}
while \(G\) is determined by
\begin{eqnarray}
&&432a^{2}c^{2}a_{1}^{2}G_{\eta}^{3} + (81a_{1}^{6} - 64a^{4}c^{2}G^{4} - 576a^{2}c^{2}a_{1}^{2}GG_{\eta\eta} - 108aca_{1}^{4}\eta G)G_{\eta}\nonumber\\
 && \qquad + 144a^{2}c^{2}a_{1}^{2}G^{2}G_{\eta\eta\eta} - 108aca_{1}^{2}G^{2}(ac\eta G - a_{1}^{2}) = 0.
\label{eq:G_final}
\end{eqnarray}

By means of the transformation,
\begin{equation}
G(\eta) = \frac{A}{P(\zeta)}, \quad \zeta = B\eta
\label{eq:final_transformation}
\end{equation}
with the scale constants \(A\) and \(B\) being determined by
\begin{equation}
B = -2^{-3/4}3^{-1/4}c^{-1/2}\beta^{-1/4}a_{1}^{1/2}, \quad A = -\frac{3a_{1}B\sqrt{-\beta}}{\sqrt{2}a}
\end{equation}
and denoting \(a_{1}\) as
\begin{equation}
a_{1} = \frac{3c\sqrt{-\beta}}{\sqrt{2}b_{1}}
\end{equation}
the \(G\)-equation (\ref{eq:G_final}) can be solved by an extended Painlev\'e IV equation
\begin{equation}
P_{\zeta\zeta} = \frac{1}{2}\frac{P_{\zeta}^{2}}{P} + \frac{3}{2}P^{3} + 4b_{1}\zeta P^{2} + 2(b_{1}^{2}\zeta^{2} - \alpha)P + \frac{\beta}{P}.
\label{eq:extended_PIV}
\end{equation}

The extended Painlev\'e IV equation (\ref{eq:extended_PIV}) becomes the standard Painlev\'e IV equation only for \(b_{1} = 1\) and \(b_{1}\) can not be scaled out for \(\zeta\) and \(P\). Thus, the solution (\ref{eq:bouss_sol2_u}) denotes the extended Painlev\'e IV soliton for $b_1\neq1$ and the Painlev\'e IV soliton for $b_1=1$.

This work introduces and systematically constructs a new class of exact solutions termed {Painlev\'e solitons}, which emerge from the nonlinear interplay between classical solitons and Painlev\'e wave backgrounds in integrable systems. By leveraging {nonlocal residual symmetries} and a novel {symmetry decomposition method}, we have explicitly derived:

\begin{itemize}
    \item \textbf{(Extended) Painlev\'e II solitons} for the KdV equation, and
    \item \textbf{(Extended) Painlev\'e IV solitons} for the Boussinesq equation.
\end{itemize}

These solutions represent a natural yet profound generalization of the well-known {elliptic solitons}, where the periodic cnoidal wave background is replaced by a non-periodic, transcendentally rich Painlev\'e wave. The concept of Painlev\'e solitons thus bridges two central themes in integrable systems: the robust, localized structure of solitons and the complex, asymptotic behavior of Painlev\'e transcendents.

The methodology employed here---{residual symmetry reduction}---proves to be a powerful and unifying framework. It allows for the decomposition of the full system into consistent dynamical subsystems in both temporal and spatial variables, facilitating the explicit construction of hybrid solutions. This approach not only recovers known elliptic solitons but also leads to new types of reductions, including the {extended Painlev\'e II equation} (\ref{eq:extended_PII}) and the {extended Painlev\'e IV equation} (\ref{eq:extended_PIV}), which generalize their classical counterparts and admit richer solution structures.

From a physical perspective, Painlev\'e solitons are expected to describe wave phenomena in settings where a coherent solitary wave interacts with a non-uniform, non-periodic background---such as in turbulent or inhomogeneous media. Their mathematical structure, combining localization and transcendental complexity, may offer new insights into the long-time asymptotics of integrable systems and the transition from order to chaos.

Several open questions and future directions arise from this work:

\begin{itemize}
    \item Can Painlev\'e solitons be observed experimentally or numerically in physical systems described by the KdV or Boussinesq equations?
    \item How do these solutions behave under perturbation or in nearly-integrable settings?
    \item Can the symmetry decomposition method be applied to other integrable systems (e.g., KP and nonlinear Schr\"odinger) to generate new Painlev\'e-type solitons?
    \item What is the role of Painlev\'e solitons in the context of inverse scattering, Riemann--Hilbert problems, Darboux transformation and Hirota's bilinear algorithm?
\end{itemize}

In conclusion, this study not only enriches the taxonomy of exact solutions in integrable systems but also demonstrates the continuing relevance of symmetry-based methods in uncovering new mathematical structures. The concept of Painlev\'e solitons opens a promising avenue for future research at the intersection of soliton theory, Painlev\'e analysis, applied mathematics, and physics.

\section*{Acknowledgement}
The work was sponsored by the National Natural Science Foundations of China (Nos.12235007, 12001424, 12271324,12501333), the Natural Science Basic research program of Shaanxi Province (No. 2021JZ-21, 2024JC-YBQN-0069), the China Postdoctoral Science Foundation (2020M673332, 2024M751921), the Fundamental Research Funds for the Central Universities (GK202304028), and the 2023 Shaanxi Province Postdoctoral Research Project (2023BSHEDZZ186), Xi'an University, Xi'an Science and Technology Plan Wutongshu Technology Transfer Action Innovation Team 25WTZD07.
The authors are indebted to thank Profs. M. Jia, Q. P. Liu, X. B. Hu, B. F. Feng and X. Z. Hao for their helpful discussions.

\bibliographystyle{plain}

\begin{thebibliography}{99}

\bibitem{zakharov1971} Zakharov V E and Shabat A B 1971 Zh. Eksp. Teor. Fiz. \textbf{61} 118

\bibitem{ablowitz1991} Ablowitz M J and Clarkson P A 1991, Solitons, nonlinear evolution equations and inverse scattering. Cambridge University Press.

\bibitem{ablowitz1980} Ablowitz M J Ramani A and Segur H 1980 J. Math. Phys. \textbf{21} 715

\bibitem{weiss1983} Weiss J, Tabor M and Carnevale G 1983 J. Math. Phys. \textbf{24} 522

\bibitem{clarkson2003} Clarkson P A 2003 J. Comput. Appl. Math. \textbf{153} 127

\bibitem{smirnov2002} Smirnov A O 2002 CRM Proceedings and Lecture Notes \textbf{32} 287, arXiv:Math/0109149

\bibitem{ling2023} Ling L M and Sun X 2023 Stud. Appl. Math. \textbf{150} 135

\bibitem{nijhoff2023} Nijhoff F W, Sun Y Y and Zhang D J 2023 Commun. Math. Phys. \textbf{399} 599

\bibitem{Dickey2003} Dickey L A 2003 Soliton equations and Hamiltonian systems. Second edition. Advanced Series in Mathematical Physics, 26, Singapore: World Scientific.

\bibitem{li2022} Li X and Zhang D J 2022 J. Nonlinear Sci., \textbf{32} 70

\bibitem{li2025} Li X, Sun Y Y and Zhang D J 2025 Nonlinearity \textbf{38} 105024

\bibitem{olver1993} Olver P J 1993, Applications of Lie groups to differential equations, Springer, New York.

\bibitem{bluman1989} Bluman G W and Kumei S 1989, Symmetries and differential equations, Springer-Verlag

\bibitem{lou1997} Lou S Y and Hu X B 1997 J. Math. Phys. \textbf{38} 6401

\bibitem{hu2009} Hu X B, Lou S Y and Qian X M 2009 Stud. Appl. Math. \textbf{122} 305

\bibitem{gao2013} Gao X N, Lou S Y, Tang X Y 2013 JHEP \textbf{05} 029

\bibitem{guo2012} Guo H, Xia T and Zhang Y 2012 J. Phys. A: Math. Theor. \textbf{45} 055202

\bibitem{liu2018} Liu S J, Tang X Y and Lou S Y 2018 Chin. Phys. B \textbf{6} 060201

\bibitem{cao1990} Cao C W 1990 Sci China Ser A. \textbf{33} 528

\bibitem{cheng1991} Cheng Y and Li Y S 1991 Phys Lett A. \textbf{157} 22

\bibitem{hao2022} Hao X Z and Lou S Y 2022 Math. Meth. Appl. Sci. \textbf{45} 5774

\bibitem{lou1999} Lou S Y and Chen L L 1999 J. Math. Phys. \textbf{40} 6491

\bibitem{Lou2023} Lou S Y, Hao X Z and Jia M 2023
JHEP \bf 03 \rm 018

\bibitem{Lou2025} Wu H L and Lou S Y 2025
 Chin. Phys. Lett. \bf 42 \rm 090003
\end{thebibliography}

\end{document}